# Why it takes a village to manage and share data

Christine L. Borgman*, Distinguished Research Professor, Information Studies, UCLA
Philip E. Bourne, Dean, School of Data Science, University of Virginia

**It takes a village to raise a child – African proverb.**

## Abstract

Implementation plans for the National Institutes of Health policy for data management and sharing, which takes effect in 2023, provide an opportunity to reflect on the stakeholders, infrastructures, practice, economics, and sustainability of data sharing. Responsibility for fulfilling data sharing requirements tends to fall on principal investigators, whereas it takes a village of stakeholders to construct, manage, and sustain the necessary knowledge infrastructure for disseminating data products. Individual scientists have mixed incentives, and many disincentives to share data, all of which vary by research domain, methods, resources, and other factors. Motivations and investments for data sharing also vary widely among academic institutional stakeholders such as university leadership, research computing, libraries, and individual schools and departments. Stakeholder concerns are interdependent along many dimensions, seven of which are explored: what data to share; context and credit; discovery; methods and training; intellectual property; data science programs; and international tensions. Data sharing is not a simple matter of individual practice, but one of infrastructure, institutions, and economics. Governments, funding agencies, and international science organizations all will need to invest in commons approaches for data sharing to develop into a sustainable international ecosystem.

## Keywords
Data sharing, research policy, data management, biomedicine, incentives, scientific practice

# 1 Introduction
The biomedical sciences have a long history of sharing data for purposes of advancing science and improving clinical practice. As the volume of research data in digital form exploded in the late 20[th] century, funding agencies began to formalize policies for sharing data. In 2003, the U.S. National Institutes of Health issued their "final statement" on sharing research data, following a community consensus process (National Institutes of Health, 2003). The 2003 policy, which requires grantees to share data resulting from their NIH awards and to submit data management plans, remains in effect. In 2020, the NIH released a new policy on data management and sharing, following a similar community consensus process, which takes effect in January 2023 (Office of The Director, National Institutes of Health, 2020).

The intervening 20 years between these two NIH policies for data sharing and management are characterized by vast increases in the scale of data production and in tools and services to exploit those data. While the biomedical sciences, physics, and astronomy remain leaders in data management and sharing, other fields are catching up. Data sharing, data repositories, pre-print servers, and datasets attached to articles at the time of publication are becoming the norm in much of the sciences and social sciences, with increasing adoption in the humanities (Borgman, 2015). The FAIR (Findable, Accessible, Interoperable, and Reusable) principles, promulgated in 2016 (Wilkinson et al., 2016), established an international framework for data sharing and reuse. The biomedical community is actively engaged in implementing the FAIR principles (Bahim et al., 2020; Celebi et al., 2020; European Union Publications Office, 2018; *FAIRsharing*, 2021; Feijoó et al., 2020; Goble et al., 2019; Groth et al., 2019; Hank & Bishop, 2018; Helliwell, 2019; Katz et al., 2021; Leonelli et al., 2021; Nitecki & Alter, 2021; Ochsner et al., 2017; Sansone et al., 2019; Schuler et al., 2019; Tammaro & Caselli, 2020).

Important milestones in biomedical data sharing predate the initial NIH mandate in 2003. These include the Bermuda Principles, which were established in 1996 as part of the Human Genome Project (Maxson Jones et al., 2018), to promote rapid sharing of data for all to use, and the Protein Data Bank (PDB), which recently celebrated its 50[th] anniversary. Community commitments to the PDB made depositing biological macromolecules essentially prerequisite to publication (Commission on Biological Macromolecules, 1989; *RCSB PDB*, 2021). These milestones, combined with subsequent NIH data sharing mandates, have contributed to the rich variety of biomedical data now available for public consumption. These data resources fueled the emergence of bioinformatics, biomedical informatics, systems biology, and most recently, biomedical data sciences (Bourne, 2021). As the value of data aggregation and analysis, beyond what any one researcher can accumulate, becomes more apparent, a virtuous cycle of data sharing and reuse has arisen in many areas of the biosciences.

Research practice has evolved more slowly in the 20 years between NIH policies, for several very human reasons. A researcher's reluctance to share data is not simply a matter of selfishness, as some have claimed. Community incentives for exchanging data are clear; incentives for individual scientists to invest in data management and distribution are less apparent. Disincentives abound, including lack of skills, lack of curatorial personnel, lack of infrastructure for managing and archiving, concerns about privacy and confidentiality of human subjects' records, intellectual property rights, concerns about being "scooped," misuses of data, labor to assist data reusers in interpretation, cost, and more. Many scientists view data sharing requirements as unfunded mandates, especially when resources and infrastructure capacity lag behind policy (Borgman, 2015; Holdren, 2013). Several promising new ventures are addressing these incentive and infrastructure challenges (*Global Biodata Coalition*, 2022; National Academies of Sciences, 2022).

Managing, sharing, and curating data are team activities. Myriad stakeholders play myriad roles in knowledge infrastructures, which are "robust networks of people, artifacts, and institutions that generate, share, and maintain specific knowledge about the human and natural worlds" (Edwards, 2010, p. 17). Our goal in this article is to identify the stakeholders involved in implementing data management and sharing policies such as that set forth by the NIH, examine their respective roles in the knowledge infrastructures necessary to sustain biomedical research data, describe interdependencies among these roles, and suggest paths forward for individuals and institutions.

## 2   Individual Scientists as Stakeholders

The sole scientist working at her lab bench and writing a sole-authored journal article to report her findings is a rarity in modern science. The size of collaborations and number of authors per paper have increased steadily over recent decades (Mallapaty, 2018). Collaborations consist of an evolving caste of characters as students graduate, post-docs become professors with labs of their own, and new multidisciplinary partnerships are forged to address new scientific problems. Each research grant has a principal investigator, and often several co-investigators. At any given time, a scientist may be a PI on one or more grants and a co-PI on several others. Co-investigators may be distributed around the country or around the world. While the PI takes responsibility for performing the grant work, including managing and releasing data, many individuals are engaged in data-handling.

Data management is thus one role among many for principal investigators. Incentives and disincentives for PIs to invest in managing and sharing data vary by many factors, few of which are well understood. Incentives and practices often vary by the "volume, velocity, and variety" of data – a simple and classic categorization (Laney, 2001). Scientists who produce "big data," especially high volume machine-generated output in standard formats, have technical means to manage those data products. These kinds of bioscience research, such as genomics, operate in a sophisticated community infrastructure for data exchange, analysis, deposit, stewardship, and reuse. Conversely, scientists who produce "little data" that consist of heterogeneous variables and formats may rely on hand-crafted methods to manage their data. The range of data types and numbers of variables also may influence data management. In one example from the NIH-sponsored workshop on which this special issue is based ("Changing the Culture," 2021), brain science research may model 20k to 30k variables per study (Ferguson, 2021).

Other scale factors that may influence a principal investigator's ability to manage data include the size of the laboratory and stability of funding. Larger, more established laboratories have more resources to hire data managers than do small laboratories that survive from one grant to the next, for example. Length of study is another factor. In research areas where one experiment yields one dataset and one paper, the units of data to be managed and released are clear. In areas where data from multiple studies are combined in multiple ways, management and release of data becomes more complex. Longitudinal data are particularly difficult to release, as findings may not become apparent until many years, and many grants, are invested in data production. Few researchers have incentives to release their full corpus of longitudinal data, at least until they have mined it fully.

A community's propensity for data sharing is also related to their available knowledge infrastructures for data management. Data that are easier to manage are easier to share with others and easier to reuse. In principle, all data need to be managed, but not all data need to be shared (Martone, 2021). Individual goals for data sharing range from the minimal to the maximal. Releasing data associated with journal articles at the time of publication is now a threshold for grant-funded research in many fields. Working completely in the open, releasing data continuously as they are produced, is rare. Somewhere in between are activities such as publicly registering hypotheses and protocols for a project (*Open Science Framework*, 2021)

From the perspective of individual investigators, data sharing may be a chicken-and-egg problem. If the research area benefits from aggregating data, the community may invest in knowledge infrastructures components such as data format standards, repositories, data coordinating centers, and deposit policies. That infrastructure, in turn, enables future researchers to contribute and acquire data for reuse. Conversely, in areas that lack infrastructure, which are often those characterized by small sets of hard-won, heterogenous data, possibly in a newly emerging field or methodology, sharing is extremely labor-intensive and less routine. Broad-brush policies, whether for data sharing or other scientific practices, rarely accommodate the full diversity of scientific practices in a domain.

Scientists in areas with minimal infrastructure and few norms for data sharing may face high hurdles in meeting policies such as those of NIH. Depending on the specific research context, principal investigators may be concerned about labor and skills for data management, competition, intellectual property and patent rights, misuses of data, and obligations to tutor data reusers. Researchers sometimes are reluctant to devote resources to documenting data in ways that others can exploit them without recompense to the data producers. Commercial exploitation is also a concern in some areas (Borgman, 2015). Tensions abound. In response to an article labeling those who reuse others' data as "data parasites," a "data parasite award" was established for those who exploit others' data for new science most effectively (Longo & Drazen, 2016; *The Research Parasite Awards: Celebrating Rigorous Secondary Data Analysis*, 2021).

Paradoxes in perceived benefits of data sharing also abound. Data can be useful as barter in attracting new collaborators. Some researchers are reluctant to release data on the grounds that they will lose that advantage (Hilgartner & Brandt-Rauf, 1994). Other researchers release data as a means to attract new collaborators (Pasquetto et al., 2019). Data citation was promoted initially as an incentive for investigators to get more credit for data sharing (Uhlir, 2012). A decade on, data citation has yet to mature as a valued scholarly metric. As discussed further below, data citation can serve multiple functions in scholarship, including credit, attribution, and discovery (Borgman, 2015, 2016).

# 3 Academic Institutional Stakeholders

To understand academic institutional stakeholders, one has to understand how universities and research institutions are organized and how they are financed for sustainability. Here we develop a generic model – as organizational models vary widely – to explain motivations (or lack thereof) and roles in data sharing that drive how institutions are likely to respond to the new NIH policy. Whether public or private, the university is led by a President (or Chancellor; titles vary) who is beholden to a Board. The President's immediate reports include colleagues responsible for faculty and academic affairs (often called a provost), student affairs, and operations. Research policy and oversight usually falls under the purview of the provost. A vice president for research (VPR), reporting to the provost or President/Chancellor, typically is responsible for grant administration, human subjects compliance, and various aspects of accountability for policy, regulation, and finance associated with the research enterprise. The top research universities in the U.S. may acquire $1bn or more per year in extramural funding from public and private sources, thus the VPR has a broad span of responsibility.

## 3.1   Academic Leadership

Provosts and Vice Presidents for Research tend to be far more concerned with funding and research compliance for their own institutions than with issues such as the value of data sharing to the larger community. They are usually well versed in aspects of data sharing related to cost, export control, and compliance. More nuanced issues such as governance, privacy, cost of preservation, and funders' mandates are lower on their list of concerns, or not recognized at all. Specifics such as requirements imposed by one funding agency, even one as large as NIH, are not typically well understood by top university leadership. Given the plethora of academic disciplines and funding sources under a VPR's span of control, the lack of attention to data sharing practice and policy is understandable. In short, academic leadership, while recognizing that the digital transformation of society is under way, may be poorly equipped to manage research data generated by the institution effectively.

Research data management is hardly the only data lacunae facing academic leaders. The ability of universities to manage and exploit their administrative data and their data about teaching and learning, academic personnel, scholarly production, and other internal records is a growing concern (Aspesi & Brand, 2020; Borgman, 2018, 2020; *SPARC Roadmap for Action*, 2019). In recognition of this changing landscape, some academic institutions are appointing chief data officers and providing them with mandates to leverage the data generated by their institution more strategically.

## 3.2   Research Computing

To be fair, data sharing mandates are but one of many compliance concerns of university leadership. Interdependencies abound, as explained further below. An immediate challenge facing university leadership with respect to research data is computing infrastructure. Depending upon the scale of research at a given institution, an office of research computing may report to the vice president for research. Despite the wide adoption of cloud computing in most economic sectors, universities typically support their own research computing infrastructure. The diversity of research methods, data practices, software, and computing practices across an institution can result in a complex network of locally controlled compute centers. Research data may live and die on computers in faculty offices, lacking an effective university infrastructure to sustain data products that may have long-term value (Borgman, 2019).

In universities with an academic medical center, research computing for clinical study usually operates on an infrastructure separate from the general campus. Such separation is necessary to meet privacy, confidentiality, and security standards that are specific to biomedical data, such as HIPAA, and more recently FISMA (Allen, 2021; Centers for Disease Control and Prevention, 2019; Cybersecurity & Infrastructure Security Agency, 2022). Separation also spares the general campus from imposing medical access constraints on educational, administrative, and non-biomedical research data. Considerable duplication of resources may result, however. Medical campus research computing offices are likely to take the brunt of responsibility for implementing NIH requirements for data sharing. Administrators may view this responsibility as an unfunded mandate, and assign the labor for governance, data management, and maintenance practices to principal investigators.

While it may be optimistic to assume that federal agency mandates, like those from the NIH, will drive better data management and governance models within universities, the less aggressive mandates in place for the last 20 years have yet to result in strong local governance at most institutions, other than policies for privacy and information security. That attitude may be changing as society at large, and by association academic institutions, comes to recognize the value of digital data to the success of an enterprise. Academic medical centers are at the forefront of this shift. Their needs for more effective data sharing lead to greater coordination between university computing centers, libraries, schools, and departments.

### 3.3 University Libraries

University libraries are institutional participants in data sharing to varying degrees. Some research universities are investing in local data repositories. Harvard libraries partnered with the Dataverse initiative of Institute for Quantitative Social Sciences more than 15 years ago to host and provide access to datasets from any academic discipline. Dataverse is now an international consortium with local instances in many universities and countries (Crosas, 2011; Harvard University, 2021; King, 2007). More commonly, university libraries are maintaining open access repositories that contain journal articles and other scholarly products, including some datasets. Other universities are indexing datasets produced by their community, without hosting the data, per se.

Libraries have a long history of stewarding the scholarly record, whether in analog or digital media. Major research libraries manage and preserve scholarly products on paper, papyrus, vellum, bamboo, microfilm, videotape, cassette, and other new media as they are invented. They have the necessary expertise in preservation, data curation, data management, data governance, archiving, records management, and conformance to standards. Libraries also have taken a central role in training researchers how to develop data management plans for the NIH, NSF, and other funders (Corti et al., 2014; Ray, 2014; Scott et al., 2013).

However strong a university library's research data management (RDM) expertise, they address these challenges amidst shrinking library budgets. Faculty, and hence administrative support for increasing library budgets, are frequently hampered by misconceptions that digital knowledge is 'free,' or at least cheaper than print materials. Neither is true. The costs of subscriptions, analog and digital materials, collection management systems, integrated library systems, computing, and staff continue to accelerate. University medical libraries are among the library units most heavily engaged in RDM, in training researchers in how to manage data, and in training researchers and grant writers how to develop data management plans that comply with funder policies (Borghi & Gulick, 2018, 2020).

### 3.4 Schools and Departments

University administrators tasked with implementing data sharing policies are faced with a "wild west" of practices within and between schools and departments. Little consistency is to be found across these entities as to how research data are perceived, maintained, or shared. Some units maintain repositories for their faculty and students, others rely on institutional repositories, others rely on thumb drives or removeable disks stored in individual offices or student carrels.

Some research communities and publishers encourage data sharing via public repositories such as Zenodo, Dryad, ICPSR, and Figshare. Communities often cluster around repositories with varying degrees of maturity and sustainability, an issue identified in the early days of open science (U.S. National Science Board, 2005). Similarly, communities cluster around different preprint servers to deposit articles and documents, such as arXiv, SSRN, Bepress, BioRxiv,

MedRxiv, and SocArxiv. Myriad repositories exist for particular data types. These repositories, in turn, partner with research platforms such as Github and OSF (Open Science Framework).

Research workflow software that streamlines biomedical research, integrating calls to other software tools and data resources, also is a critical part of this ecosystem (Bechhofer et al., 2013; Belhajjame et al., 2015; Clark et al., 2014; Ferreira da Silva et al., 2021; Garijo et al., 2013; Goble et al., 2010; Sansone et al., 2019). Whether the new NIH data sharing policy can bring coherence to this complex culture remains to be seen.

In summary, relationships between university stakeholders, researchers, librarians, research and administrative computing staff are insufficiently aligned with respect to data maintenance, preservation, sharing, and effective reuse. Private enterprise, in contrast, increasingly sees data products as major assets to be exploited. Several universities are notable exemplars in data utilization. Course Signals, developed at Purdue University, is used to gauge students' academic preparation, engagement, and academic performance (Arnold & Pistilli, 2012). Georgia State University has something similar, although not without controversy over student privacy (Barshay & Aslanian, 2019). Northeastern University created a virtual analytics center of excellence to promote effective use of university data (Krawitz et al., 2018). However, tensions among university stakeholders abound with respect to data exploitation, privacy, and security (Borgman, 2018).

Few stakeholder institutions are well positioned to comply with the new NIH data sharing policy. The larger practical challenges to implementation include direct appeals to university leadership to recognize research data as community assets, to address data sharing policy compliance within their institutions, to invest in infrastructure for research data management, and to promote RDM training for new generations of researchers. Such appeals will be of little use to less well-resourced research institutions, which risks further expansion of the digital divide as experienced by minority-serving institutions, community colleges, and others (Buzzetto-Hollywood et al., 2018; Jurado de los Santos et al., 2020).

# 4   Interdependencies

Biomedical researchers and universities subject to the NIH data sharing policies are the primary, but not sole, stakeholders in sharing, using, reusing, managing, and stewarding these research materials. They exist together in a complex knowledge infrastructure. As discussed in more depth earlier (Borgman, 2015), knowledge infrastructures have the economic characteristics of a 'knowledge commons,' defined most simply as a 'resource shared by a group of people that is subject to social dilemmas' (Hess & Ostrom, 2007, p. 3). Shared community resources such as research data must be governed in ways that address distribution of power, influence, authority, and funding. Technical issues also abound in governing the commons, involving centralized and decentralized resources, large silos of data, and all manner of interoperability and maintenance challenges (Denis & Pontille, 2017; Edwards, 2019; Graham & Thrift, 2007; Lane, 2020; Nosek et al., 2015; Ostrom & Ostrom, 1977; Vinsel & Russell, 2020).

Here we outline seven interdependent concerns of individual researchers, academic institutions, funding agencies, governments, industry, and other stakeholders in sharing biomedical research data. These are among the most salient interdependences in the current landscape; the list is far from complete.

## 4.1   What Data to Share

An overarching issue is the lack of agreement on what constitutes sharable data, or even what constitutes "data" (Borgman, 2015; Leonelli, 2016). A limitation of data sharing policies is the difficulty of defining precisely what are considered "research data" subject to the policy. The NIH policy definitions, as shown in Box 1, refer to "recorded factual material" and specify judgment criteria such as "commonly accepted in the scientific community" and the degree of quality expected. The definition also says that data are to be shared whether or not used to support scholarly publications. The policy offers some guidance on what are not considered data, such as laboratory notebooks, informal communications, and physical objects.

## Section II. Definitions

For the purposes of the DMS Policy, terms are defined as follows:

| | |
|---|---|
| **SCIENTIFIC DATA** | The recorded factual material commonly accepted in the scientific community as of sufficient quality to validate and replicate research findings, regardless of whether the data are used to support scholarly publications. Scientific data do not include laboratory notebooks, preliminary analyses, completed case report forms, drafts of scientific papers, plans for future research, peer reviews, communications with colleagues, or physical objects, such as laboratory specimens. |
| **DATA MANAGEMENT** | The process of validating, organizing, protecting, maintaining, and processing scientific data to ensure the accessibility, reliability, and quality of the scientific data for its users. |
| **DATA SHARING** | The act of making scientific data available for use by others (e.g., the larger research community, institutions, the broader public), for example, via an established repository. |
| **METADATA** | Data that provide additional information intended to make scientific data interpretable and reusable (e.g., date, independent sample and variable construction and description, methodology, data provenance, data transformations, any intermediate or descriptive observational variables). |
| **DATA MANAGEMENT AND SHARING PLAN (PLAN)** | A plan describing the data management, preservation, and sharing of scientific data and accompanying metadata. |

Box 1: Definitions in NIH Data Policy (Office of The Director, National Institutes of Health, 2020)

The NIH definitions, while constructive and succinct, leave plenty of room for interpretation. As Sansone discussed in a workshop panel (Sansone, 2021), policies tend to be "vague, aspirational, and flexible." They are only as effective as their implementation and subsequent conformance. Individual communities have varying standards and practices for what is "commonly accepted." Practices evolve over time, vary by type of data, and may not be formally documented. In some research areas, the "raw data" of initial observations may diverge along multiple paths as datasets are cleaned of artifacts, "reduced," and combined. Processes such as data cleaning, calibration, and modeling are not mentioned in the policy, for example. The opinions of individuals and communities about how many levels of data need to be kept or shared also diverge. The sheer practicality of data release is often a barrier to data sharing, given the wide array of guidelines for what to release, and in what form. A recent case study of datasets available on GitHub suggests some factors that may influence dataset reusability (Koesten et al., 2020).

The scope of metadata deemed necessary "to make scientific data interpretable and reusable" also varies. Some researchers may provide just enough metadata to make a dataset interpretable, such as variable names and scientific units. Others may provide extensive documentation of protocols, models, and software. Researchers who adhere strongly to open science principles may post sufficient software, tools, data, and protocols to support computational reproducibility of their projects (Brinckman et al., 2019; McNutt, 2020; Stodden, 2009, 2020). How researchers and program officers determine what constitutes data subject to sharing will determine reusability, discoverability, and resources to sustain those data. The FAIR Guiding Principles (Wilkinson et al., 2016), described above, have been widely adopted. Their implementation through efforts such as GO FAIR offer examples of how to make these principles of practical use (GO FAIR, 2020).

## 4.2   Data, Context, and Credit

Data, whether open or otherwise, is only a part of the research workflow. The basis of any given study may include annotations in the form of lab notes, software, protocols, grants initiating the work, pre-publications, publications, and even Wikipedia entries and social media. Each study builds upon prior work; no study or dataset exists in isolation. Multiple interdependencies arise as consequences of the contextual nature of research data.

Without adequate information about how data interact with the contexts in which they were generated, produced, handled, and interpreted, a dataset is merely a string of bits. The problem of making data "mobile," or interpretable outside their original context, is the subject of much research in the social studies of science (Borgman, 2015; Bowker, 2005; Knorr-Cetina, 1999; Latour, 1987; Latour & Woolgar, 1979; Lave & Wenger, 1991; Leonelli & Tempini, 2020). In biomedicine, data interpretability is embodied in ontologies, data dictionaries, and standardized structures and formats, all of which may vary by research community.

A related interdependency is the context for reuse of data. If a dataset is to be reused primarily for comparison purposes, such as calibration, basic metadata may suffice. If a researcher wishes to employ a dataset in a novel inquiry, the prospective data reuser may need to contact, or collaborate with, the original producer, a situation termed "the data creators' advantage" (Pasquetto et al., 2019). In an era of "big data" and machine learning, in which large quantities of research data are mined to ask new questions and build new models, questions of data context and quality become even more profound (Meng, 2021).

A third interdependency between data and context is the unequal value placed on components of the scholarly record. Despite numerous calls to consider datasets first-class scholarly objects, peer-reviewed journal articles remain the coin of the realm in most scientific disciplines (Bierer et al., 2017; Fenner et al., 2019a). Tenure is unlikely to be awarded based on contributions to Wikipedia, YouTube videos, blogs – or to datasets. For these reasons, scholars spend much more time polishing their publications than polishing their datasets. Similarly, scholars are much more concerned with receiving citations to their publications than to their datasets. The slow adoption of standardized data citation methods is at least partly attributable to the low status of datasets as scholarly objects (Borgman, 2015).

A dataset may be accessed many more times than the scientific paper "advertising" that data, yet it is the paper that accrues credit to the authors. While this situation persists, incentives to provide quality data to the community will continue to lag. The scholarly system is skewed towards an era when only the journal article, conference paper, or other text product was easily available (Borgman, 2007). Antiquated metrics for scholarly value persist to the point that researchers get credit for datasets only if they write papers about those data. As a consequence, journal articles and papers acquire metadata and description as independent scholarly objects, but datasets and related research products such as software rarely acquire enough metadata to be adequately represented in the citation network. Dataset description is improving with resources like Dryad and DataCite, which support extensive metadata (Brase et al., 2015; *DataCite Schema*, 2022; *Dryad*, 2022). Similarly, schema.org provides incentives to provide rich metadata by returning structured outputs of value to the data consumer (Schema.org, 2022).

## 4.3   Data and Discovery

If a dataset is shared, but cannot be discovered by others, it might as well not exist. Few journals require authors to provide formal bibliographic citations to datasets, even when they require data deposit as a condition of publishing articles. A large, international, multidisciplinary study of data discovery methods revealed the many challenges facing researchers in locating datasets. Study participants, which included researchers, students, librarians, and other scientific professionals, reported a diverse array of strategies to search journal literature, data repositories, laboratory websites, social media, and other sources. Biomedical fields had high rates of data sharing and data searching in the study (Gregory et al., 2020). As Goble (2021) reported in the workshop, biomedical scientists structure data across multiple data stores and use multiple repositories. No unified cataloging system exists to locate data or to associate protocols with datasets.

Where de facto repositories exist for specific types of biomedical data, authors and publishers tend to cluster around them (Sansone et al., 2019). As biomedical specialties become more complex and generate more types of data, researchers may develop new repositories to share these data. As a consequence, new de facto repositories emerge as silos that do not interoperate with other resources, further limiting the integration necessary to move the overall field forward. This phenomenon is manifest in the annual database issue of *Nucleic Acids Research*, which is the 'go to' place to announce a bioscience data resource. The number of papers on such resources continues to grow. Of particular concern for the community is the flux in data resources, as new databases appear and others cease to exist (Rigden & Fernández, 2021).

Part of the challenge in discovering datasets is the evolving relationship between journals and data repositories. As proposed nearly two decades ago, journal articles and datasets could serve as mutual search mechanisms if their structures were more closely aligned (Bourne, 2005). Common systems of metadata, automatic generation of terms, and automatic text recognition of structures and datasets are well within the realm of technical possibility, but adoption lags. The value of data citation for data discovery is under-appreciated. If adequate citation linkages existed, information seekers could navigate from a paper, to a dataset, to all papers that used that dataset. As standards for data citation are adopted by journals, repositories, and authors, datasets should become more readily discoverable (Cousijn et al., 2018; Fenner et al., 2019b; Gregory et al., 2020; Lane, 2020; Lowenberg, Under review; Lowenberg et al., 2019).

## 4.4   Data Assets as Research Methods

Another interdependency between stakeholders that arose several times in the workshop is the locus of learning how to manage and share data. Graduate courses in research methods is the usual answer to this question, at least in the social sciences. By embedding data management instruction in research design, the thinking goes, students will learn to treat their research data as assets to be managed, shared, discovered, and reused (Borgman, 2015). As Wyatt (2021) and others reported at the workshop, few areas of the biosciences teach formal graduate courses in research methods. In fields where research design is learned through mentorship, fewer mechanisms exist to develop standards, practices, and infrastructures necessary for data sharing.

## 4.5   Intellectual Property in Data

Legal aspects of data sharing in the biosciences were addressed in several workshop sessions and covered more fully in other articles in this special issue. One legal issue to highlight in the interdependencies between stakeholders involves intellectual property and licensing. For example, data sharing mandates can conflict with licensing rights. The Bayh-Dole Act of 1980 stipulates that universities can retain the rights to their inventions (*The Bayh-Dole Act: A Guide to the Law and Implementing Regulations*, 1999). However, patentable inventions may involve data that are supposed to be shared under the conditions of the grant award. Principal investigators cannot simultaneously protect those data and make them publicly available under an open license.

Complicating matters further are the challenges of identifying the few inventions that will have great financial value – separating the wheat from the chaff, as it were. In biomedicine, some discoveries do lead to blockbuster drugs or medical devices that are financially lucrative. Legendary lawsuits between universities sometimes result (Isaacson, 2021). Within universities, intellectual property rights for research products usually are negotiated by technology transfer offices that report to the office of research, or directly to the provost or president. Because tech transfer offices are seeking potential "cash cows," they may have a misplaced sense of value of intellectual property to the overall research enterprise. Many useful, but monetarily worthless, innovations get buried in university tech transfer offices for months or years. Data are caught up in this corundum if they are associated with such an invention. Arguments for the community value of releasing these data get little traction when money is at stake.

Universities, funding agencies, and other stakeholders are exploring approaches to resolving these conflicts (Grabus & Greenberg, 2019; Stodden, 2009). Some of these conflicts can be addressed as temporal matters, whether by imposing short deadlines on tech transfer decisions or by establishing clear proprietary periods during which researchers can control their data prior to release.

Other approaches involve licensing, such as Creative Commons licenses favored by open science advocates or fee-based licenses associated with patents or inventions. Technology issues arise here too, such as the lack of licensing metadata on datasets (Moore et al., 2016). Image repositories such as *Unsplash*, *Flickr*, and *Google images*, for example, label individual objects with license rights, at least where known. Image-seekers can specify licenses, among other search parameters. Retrieved images contain attribution and rights information.

Lastly, and briefly, are the complex relationships between universities and the biotech industry. Industry benefits by incorporating open data in products that are proprietary, but they are not obligated to release data from privately funded research. Indeed, industry has little incentive to make high quality data available to the academic community and has no business model to do so. Where industry benefits from open data, investigators may find disincentives to share. Some important industry-academic-open science initiatives have found ways to create, share, and use open data for research and for development (Williamson, 2000). These partnerships include *Sage Bionetworks*, the *Structural Genomics Consortium*, *Accelerating Medicine Partnership*, and federal grant programs for public-private partnerships.

## 4.6 Data Science Initiatives

In the U.S., data science emerged as an academic field around 2010 through the Data Science Environments Partnership supported by the Gordon and Betty Moore and Alfred P. Sloan Foundations, which in turn led to the Academic Data Science Alliance (Academic Data Science Alliance, 2022). Hundreds of institutional data science initiatives now exist in the U.S. alone, whether as centers, departments, schools, divisions, or cross-campus initiatives. These data science programs typically juxtapose statistics, computer science, and mathematics as combined approaches to every imaginable vertical means of generating digital data. A growing number of data science programs include information studies, social sciences, humanities, and philosophy as participating departments.

Bioinformatics emerged as a data science specialty in biomedicine in response to the human genome project (Watson, 1990). Even in biomedicine, something more profound is happening, aptly named 'biomedical data sciences' and characterized by a broader array of data types and more complex methods of analysis (Bourne, 2021). Taken together, they represent the long-sought concept known as 'multiscale modeling' – from molecules to populations.

The need to implement multiscale modeling was amplified by the need to address the COVID-19 pandemic. As a global crisis, COVID has led to an unprecedented degree of data sharing. COVID-related science encompasses data at all biological scales, from deidentified data on large cohorts of patients derived from the electronic health record, to imaging data on organs, to biopsies on tissues, to genetic sequencing at the molecular level (Meng, 2020). The challenges are to use the genotype to explain the phenotype, and to design appropriate interventions that lead to positive health outcomes. Just one example, taken from our own work, is the observation that women at different ages are affected differently than men in terms of both COVID disease severity and death. That difference can be attributed to the biochemical role of estradiol (Seeland et al., 2020)

University libraries, as data suppliers, and data science initiatives, as data consumers, are establishing productive partnerships. Opportunities arise for synergy across the education, research, and service missions of the university. University libraries are taking the lead in offering courses in using, managing, and disseminating data products, and in providing institutional repositories for research outputs. Lastly, institutional governance issues faced by data scientists are being addressed with the aid of librarians who understand copyright law and are versed in the work of open knowledge communities.

## 4.7 Thinking Globally, Acting Locally

Biomedical data are of international value and typically made available globally, yet most are funded and maintained nationally or locally. Scientifically advanced nations, notably the US and Europe, established a quid pro quo for data sharing early in the digital age (*Bits of Power*, 1997; Collins et al., 2003). These countries continue to invest in knowledge infrastructures to share research data internationally, including technical and human components. As more countries seek access to these data, and international tensions arise, the model becomes strained, leading to inequalities in the global data ecosystem. Problems of governing the commons, such as 'free riders,' are well known in law, economics, and biosciences (Hess & Ostrom, 2007; Lessig, 2001; Ostrom & Ostrom, 1977; Reichman et al., 2016; Schofield et al., 2010).

International tensions in data access are compounded by differences in privacy laws and policies for data sharing. Complicating matters further are technology, law, and policy related to transborder data flows. Laws may vary by the location where data were collected, where they are held, and other jurisdictional matters. Global bodies such as the International Science Council, which represents national and regional science organizations, are the usual means for coordinating activities such as data sharing. CODATA, for example, coordinates scientific data activities through national science academies as part of the International Science Council. Non-governmental groups such as the Research Data Alliance also aid in coordination and educational efforts (Berman & Crosas, 2020; Lide & Wood, 2010).

The Protein Data Bank, with its 50-year history, is a useful example of international cooperation. Identical corpora of PDB raw data are maintained on three continents – North America, Asia, and Europe – at significant cost and duplication of effort. The motivation for such duplication is that each region wishes to be identified with biological data that has significant local impact. An advantage of the duplicate corpora is the ability to provide tailored services layered on top of the raw data. Moreover, scientists associated with the three data sources agree on common standards, consistency of the raw data, and policies for data sharing across continents.

Tensions between local and global priorities in data sharing exist at many levels. Within countries, individual funding agencies have their own policies, even within biomedicine. The NIH data sharing policies are not fully aligned with those of the Howard Hughes Medical Institute or the Gates Foundation, for example, despite common goals. Attitudes, policies, and capabilities for data sharing vary widely within and between individual universities. These differences may be driven by factors such as private versus public institutions, degree of research focus, depth of endowment, and other aspects that are difficult to assess.

## 5   Building the Village

Data sharing is not merely an act of releasing a digital product to the world. Rather, data sharing has all the emergent properties of a complex system. Many stakeholders, knowledge infrastructures, motivations, practices, and policies interact in unpredictable ways. Implementing the NIH data sharing policy is best addressed as a problem of the commons, sometimes also known as a 'collective action problem' in which individual actors pursue their own best interests over the long-term interests of the community.

The long list of challenges identified in this article is not exhaustive, nor are these factors mutually exclusive. Attempts to address them individually are unlikely to achieve global solutions. Rather, a starting point is to acknowledge the complex nature of data sharing, and to see it as a collective challenge rather than solely an individual responsibility of principal investigators. While PIs play essential roles in producing sharable data resources, they can execute those roles only in concert with many other stakeholders.

Promising actions to construct, strengthen, and sustain the necessary knowledge infrastructures for robust dissemination of research data include distributing responsibility among multiple actors, investing in data management expertise, and reframing goals in collective terms. Overall, a much more holistic approach is required than laying full responsibility for data sharing on the shoulders of principal investigators. Investing in data science, developing a cadre of data curators, and deploying those data professionals throughout the research enterprise all will be steps forward.

Given the commons nature of data exchange, funding agencies will need to support data repositories for the foreseeable future. Scaling up data repositories to the rate of data production in the biosciences is a daunting challenge for policy, technology, and economics. Far more work is needed to develop criteria for what data are worth keeping, in what forms, for how long, and by whom.

Signs of a more holistic approach are on the horizon. Funding agencies and academic institutions are at the forefront of progress improving the ecosystem of resources, incentives, and responsibilities. The Global Biodata Consortium is working towards the more equitable and coordinated funding of open data resources (*Global Biodata Coalition*, 2022). Likewise, the Higher Education Leadership Initiative for Open Science (HELIOS) convened by the National Academies of Science Engineering and Medicine (NASEM) has over 50 institutions signed up to exchange best practices on institutional open knowledge practices, including data sharing and reuse (National Academies of Sciences, 2021; *Roundtable on Aligning Incentives for Open Science | National Academies*, 2022).

Returning to the village metaphor, a village typically has elders (the funders, NIH in this case), marketplaces, governing bodies, schools, points of aggregation (academic institutions), a population (individual researchers), and relationships to those outside of the village (the general public). All must work in unison if the village is to prosper. The elders will be followed by the people only if what they say makes sense and the people can see that a strategy is advantageous. The new NIH data sharing policy, effective in 2023, appears to make sense to many stakeholders. Will 'the people' follow the policy? Yes, if encouraged by the thought-leaders and by the agencies that provide support. Village institutions, like schools and marketplaces, prosper if the people prosper – in this case though a greater number of successful grants. Villages have existed and prospered for the past 12,000 years, offering hope that communities will find sustainable paths for data sharing in turn.

# 6   Acknowledgements

We are grateful to Jessica Levinson, a professor at Loyola Law School; Susanna-Assunta Sansone, a professor at the University of Oxford; and two anonymous reviewers, for their thoughtful comments on earlier drafts of this article. We thank Cam Mura, a senior scientist at the University of Virginia, and Megan Haas, a student at James Madison University, for creating the Data Village diagram that accompanies this article.